\documentclass{PoS}
\usepackage{wrapfig}
\usepackage{amsmath,amssymb}

\newcommand\GeV{\mathrm{GeV}}
\newcommand\TeV{\mathrm{TeV}}
\newcommand\nb{\mathrm{nb}}
\newcommand\pb{\mathrm{pb}}

\title{Event shapes at hadron colliders}

\ShortTitle{Event shapes at hadron colliders}

\author{\speaker{Andrea Banfi}
  \thanks{This work has been done in collaboration with Gavin Salam (LPTHE Jussieu, Paris) and Giulia Zanderighi (Oxford University).}\\
  ETH Zurich, Institute for Theoretical Physics\\
  E-mail: \email{banfi@itp.phys.ethz.ch}}

\abstract{We present precision results for distributions in global
  event shapes that can be measured at hadron colliders within
  experimental limitations. These predictions are obtained by
  combining exact next-to-leading order (NLO) with the all-order
  resummation of large logarithms of soft and collinear origin. We
  then discuss how event-shape measurements can be used for the tuning
  of Monte Carlo event generators, for tests of models of
  hadronisation and underlying event, and as discriminatory tools
  between QCD jet-like and New Physics events.}

\FullConference{
 DIS 2010, April 19-23, 2010, 
		Firenze, Italy \\
              Physics at the LHC 2010, June 7-12, 2010, DESY, Hamburg, Germany}

\begin{document}

\section{Introduction}
\label{sec:intro}

Event-shape variables are infrared and collinear safe measures of the
geometrical properties of the hadronic energy momentum flow, giving an
idea on whether an event is pencil-like, planar, spherical,
etc. Measurements of their mean values and distributions have played a
crucial role at LEP, for precise determinations of the strong coupling
$\alpha_s$, for tests of analytical models of hadronisation
corrections, and for validation of Monte Carlo (MC) event generators
(see~\cite{Dasgupta:2003iq} for a recent review). In spite of the
success of these studies in $e^+e^-$ annihilation, very little
attention has been devoted to their counterparts at hadron
colliders~\cite{CDF-Broadening,D0Thrust}. This was mainly because the
quantities that were conveniently measured experimentally could not be
accurately computed in perturbative QCD. Here I would like to present
event shapes that can be measured at actual hadron colliders, and
whose distributions can be computed in perturbative QCD at the
accuracy needed to have a reliable estimate of the associated
theoretical uncertainties.

\section{Event-shape variables at hadron colliders}
\label{sec:evshp-def}

In hadron-hadron collisions, we consider events with two hard central
jets, and define event shapes that vanish in the limit of two narrow
jets. For instance, given all hadrons $\{q_i\}$ in a rapidity region
${\cal C}$ (for instance $|\eta_i|<\eta_{\cal C}$) and using
transverse momenta $\vec q_{\perp,i}$ only, we define the transverse
thrust $T_\perp$ and the thrust minor $T_m$ as follows
\begin{equation}
  \label{eq:t-tmin}
  T_\perp = \max_{\vec n_\perp} 
  \frac{\sum_{i\in{\cal C}} |\vec q_{\perp,i} \cdot \vec n_\perp|}
  {\sum_{i\in{\cal C}} |\vec q_{\perp,i}|}\,,
  \qquad\qquad
  T_m = 
  \frac{\sum_{i\in{\cal C}} |\vec q_{\perp,i} \times \vec n_\perp|}
  {\sum_{i\in{\cal C}} |\vec q_{\perp,i}|}\,.
\end{equation}
We can also introduce boost-invariant event shapes involving
longitudinal degrees of freedom, like invariant masses or broadenings,
or the three-jet resolution parameter $y_3$. An extensive list of
hadronic event-shape definitions can be found
in~\cite{Banfi:2004nk,Banfi:2010xy}.

There are three basic reasons why dijet event shapes can be studied
experimentally with very first data. First, cross sections for dijet
production are large both at the Tevatron and at the LHC, as shown in
table~\ref{tab:xsct}. From table~\ref{tab:xsct} one can also see that
the flavour content of a sample can be varied by changing the leading
jet $p_t$-cut. Low-$p_t$ samples (Tevatron with $p_{t1}>50\GeV$, LHC
with $p_{t1}>200\GeV$) are gluon dominated, while initial-state quarks
become more important for high-$p_t$ samples (Tevatron with
$p_{t1}>200\GeV$, LHC with $p_{t1}>1\TeV$).
\begin{table}[h]
  \centering
{\small
  \begin{tabular}{|l|cc|ccc|}
    \hline 
    & LO & NLO & $qq\!\to\!qq$ & $qg\!\to\!qg$ & $gg\!\to\!gg$\\\hline
    Tevatron, $p_{t1} > 50\GeV$  & $60 \nb$ & $116 \nb$ 
    & 10\% & 43\%& 45\%
    \\
    Tevatron, $p_{t1} > 200\GeV$  & $59 \pb$ & $101 \pb$
    & 41\% & 43\% & 12\%
    \\
    $14\TeV$ LHC, $p_{t1} > 200\GeV$  & $13.3 \nb$ & 
    $23.8 \nb$
    & 7\% & 40\% & 50\%
    \\
    $14\TeV$ LHC, $p_{t1} > 1\TeV$  & $6.4 \pb$ & 
    $10.5 \pb$
    & 31\% & 51\% & 17\% \\
    \hline
  \end{tabular}
}
  \caption{Cross sections for the production of two jets in a central rapidity 
    region ($|y_{\mathrm{jets}}|<0.7$ at the Tevatron and $|y_{\mathrm{jets}}|<1$ 
    at the LHC) with a cut on $p_{t1}$, the transverse momentum of the leading 
    jet. On the right it is possible to see the relative importance of each 
    partonic subprocess.}
  \label{tab:xsct}
\end{table}
Second, event shapes are normalised quantities: experimental
uncertainties associated with jet energy scale cancel between
numerators and denominators, see eq.~(\ref{eq:t-tmin}). Finally, since one
usually measures normalised differential distributions, like $1/\sigma
\,d\sigma/dT_m$, no determination of luminosity is required.

\begin{wrapfigure}{l}{0.5\linewidth}
  \centering
  \includegraphics[width=.5\textwidth]{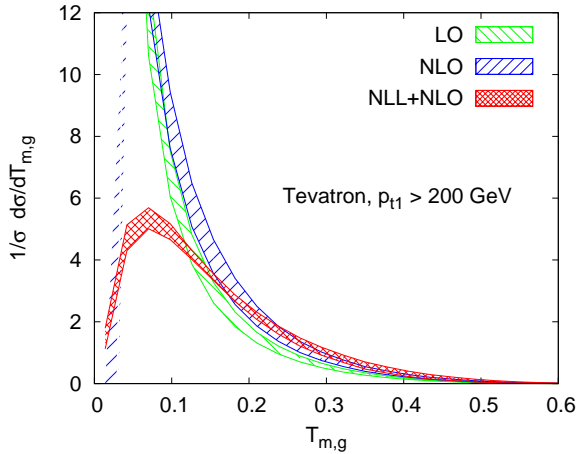}
  \caption{The differential distribution in $T_{m,g}$ for a Tevatron
    high-$p_t$ sample. The bands represent theoretical uncertainties.}
  \label{fig:tmin-nll-v-nlo}
\end{wrapfigure}

From a theoretical point of view, event-shape distributions can be
computed at NLO with \textsc{nlojet++}~\cite{Nagy:2003tz}. However, as
can be seen in the example plots of figure~\ref{fig:tmin-nll-v-nlo},
both LO and NLO predictions diverge at small values of $T_{m,g}$ (the
directly global variant of $T_m$, see
section~\ref{sec:evshp-theory}). Only a combination of NLO and
next-to-leading logarithmic (NLL) resummation (labelled NLL+NLO) gives
a distribution that is sensible for any value of $T_{m,g}$. In
particular, resummation restores the correct physical behaviour at
$T_{m,g}\to 0$, corresponding to vanishing probability of having
accelerated charges without accompanying radiation.

NLL resummation involves writing the {\em integrated} $T_{m,g}$
distribution as an exponent $\Sigma(T_{m,g})=\exp[L g_1(\alpha_s L)
+g_2(\alpha_s L)]$, with $L=\ln (1/T_{m,g})$, $g_1(\alpha_s L)$
resumming the leading logarithms (LL, $\alpha_s^n L^{n+1}$), and
$g_2(\alpha_s L)$ the NLL, $\alpha_s^n L^n$. Knowledge of
$g_1(\alpha_s L)$ determines the position of the peak of the
differential distribution, typically in the region $\alpha_s L \sim
1$. In the peak region $g_2(\alpha_s L)$ becomes of order one, and is
therefore needed to stabilise both the position and the height of the
peak.

NLL resummability is guaranteed for variables satisfying the following
conditions~\cite{Banfi:2004yd}: a specific functional dependence on a
single soft and collinear emission; (continuous) globalness, i.e.\
sensitivity to emissions in the whole of the phase space; recursive
infrared and collinear (rIRC) safety, a subtle mathematical condition
on the event-shape scaling properties with multiple emissions. If
these conditions are satisfied, the relevant emissions that contribute
to event-shape distributions at NLL accuracy are soft and collinear
parton clusters widely separated in rapidity, which, due to QCD
coherence, can be considered as independently emitted from hard
legs. Since a similar pattern of emissions is simulated by MC event
generators, one may expect that most features of rIRC safe global
event-shape distributions are correctly described by these theoretical
tools.

Of the three constraints, the most difficult to satisfy experimentally
is globalness, due to the fact that the measurement region ${\cal C}$
is preferably restricted to the central detector region (e.g.\
$|\eta|\lesssim 2.5$ at the LHC), and in any case no measurement is
actually performed in the very forward regions (corresponding to a
limiting rapidity $\eta_c= 5$ at the LHC). However~\cite{Banfi:2004nk},
one can devise classes of global event shapes even at hadron
colliders: {\em directly global}, where the region ${\cal C}$ extends
up to the maximum available rapidity $\eta_c$; {\em exponentially
  suppressed}, where ${\cal C}$ is inside the acceptance of the
central detectors (e.g.\ $|\eta|<1$ at the Tevatron and $|\eta|<1.5$
at the LHC), while outside this region we add to the event-shape
definition a term that exponentially suppresses the contribution of
hadrons in the forward regions; {\em recoil}, where measurements are
performed only in a central region $\cal C$, and we add a term that is
sensitive to emissions outside $\cal C$ through recoil. In the last
case however a numerical breakdown of NLL resummation occurs in the
region where the event shape is small.

\section{Event-shape distributions the Tevatron and at the LHC} 
\label{sec:evshp-theory}

\begin{wrapfigure}{r}{0.47\linewidth}
  \centering
 \includegraphics[width=.45\textwidth]{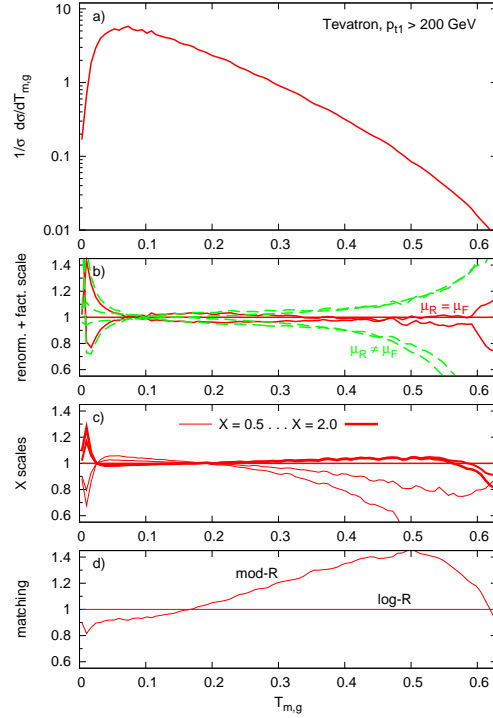}
 \caption{NLL+NLO predictions and theoretical uncertainties for the
   distribution in $T_{m,g}$.}
  \label{fig:tmin+uncerts}
\end{wrapfigure}
In~\cite{Banfi:2010xy} we have performed a NLL+NLO resummation for a
number of selected event shapes.  Figure~\ref{fig:tmin+uncerts} shows
predictions for the directly global thrust minor $T_{m,g}$, together
with theoretical uncertainties.  The latter, aimed at having an
indication of missing NNLO and NNLL corrections, are estimated via:
asymmetric variation of renormalisation and factorisation scales
$\mu_R$ and $\mu_F$; variation of the logarithm to be resummed $\ln (X
T_{m,g})$ with $1/2<X<2$; variation of the matching scheme (log-R or
mod-R).  We observe that uncertainties are under control and within
$\pm 20 \%$ in a wide range of values of $T_{m,g}$. Similar results
are obtained for all considered event shapes.

\begin{figure}[h]
  \centering
  \includegraphics[width=.45\textwidth]{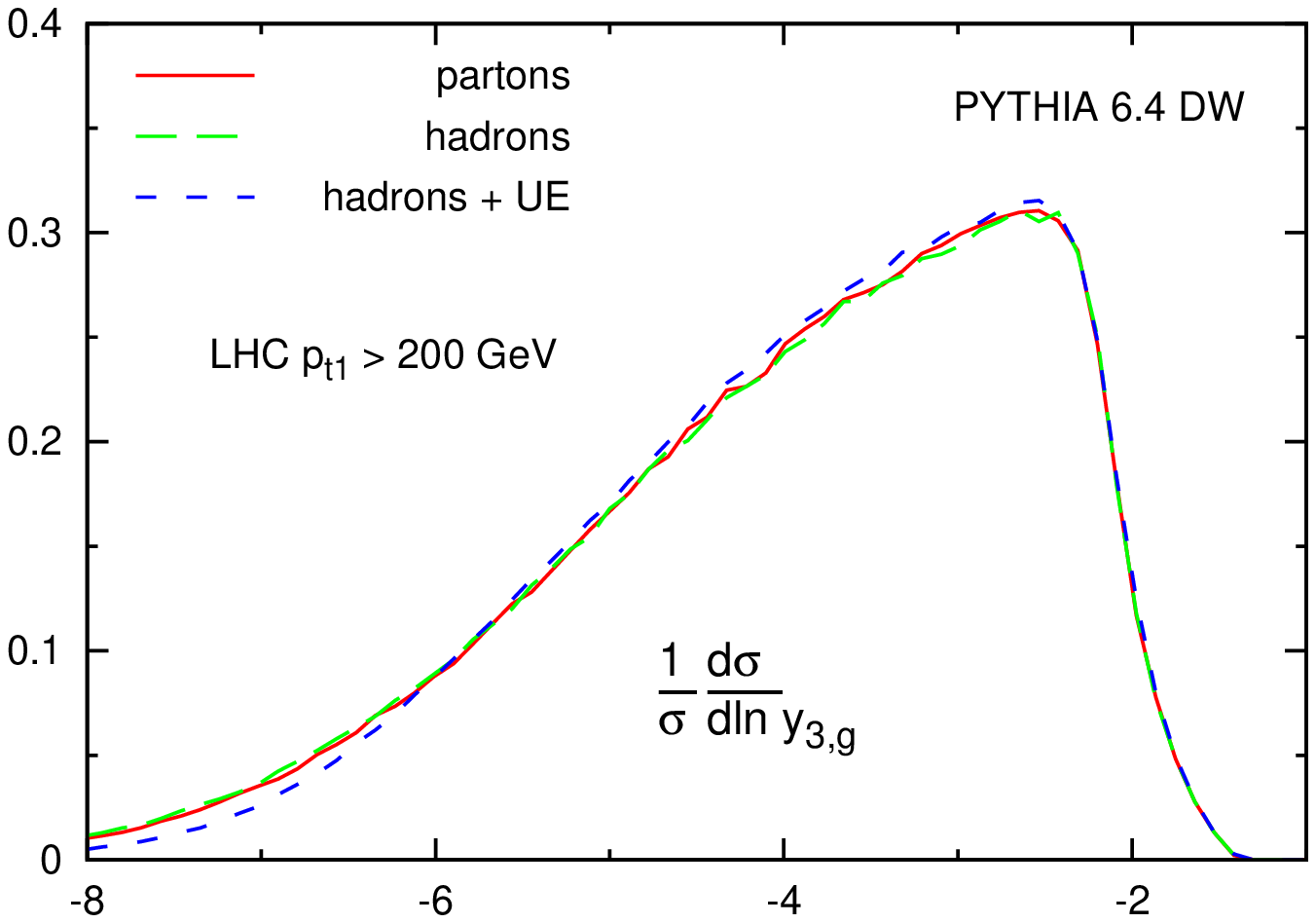}    
  \rule{.025\textwidth}{0cm}
  \includegraphics[width=.45\textwidth]{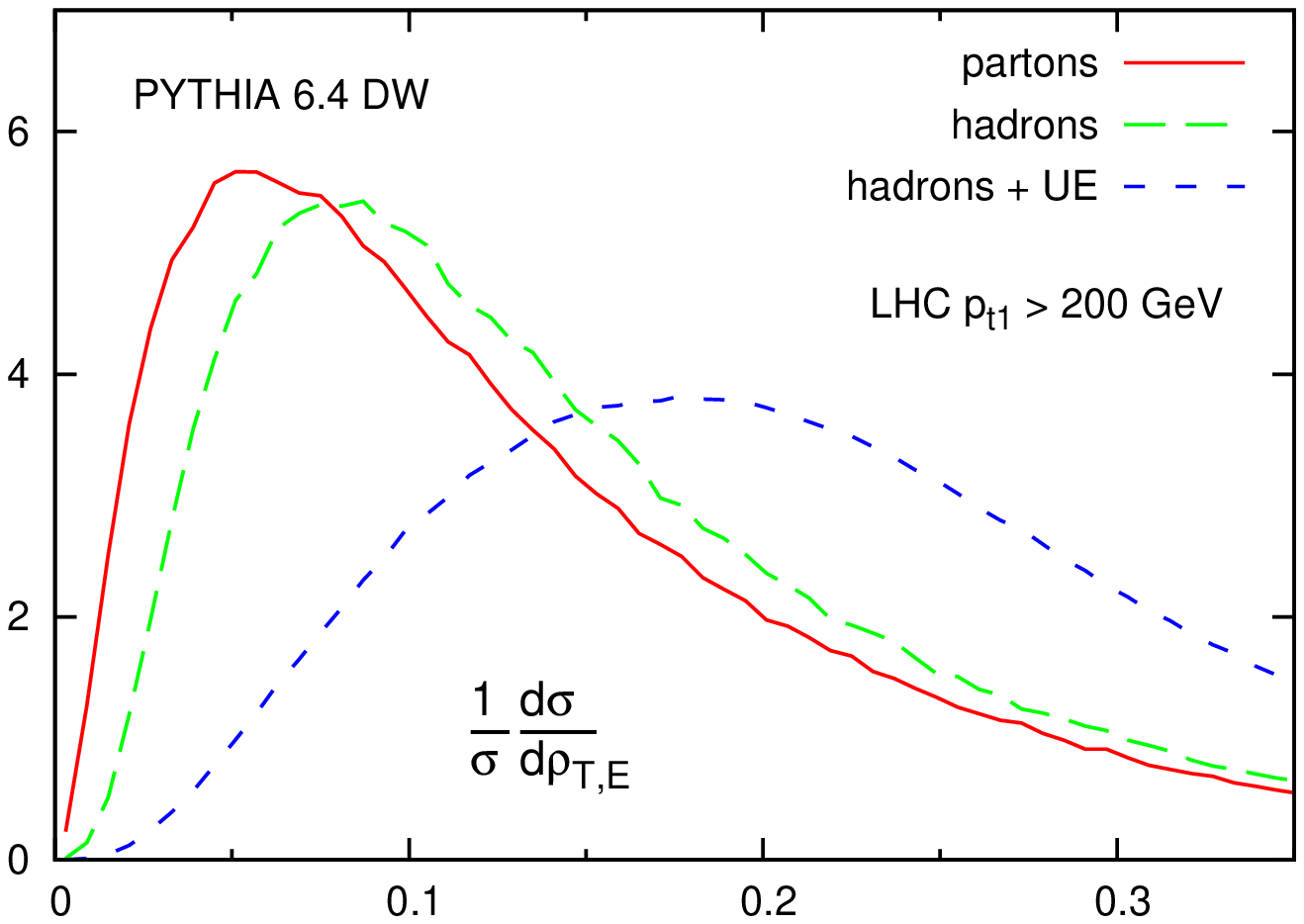}
  \caption{Distributions in directly global three-jet resolution
    $y_{3,g}$ (left) and exponentially suppressed total mass
    $\rho_{T,{\cal E}}$ (right), as obtained with the MC event
    generator \textsc{pythia}~\cite{Sjostrand:2006za} for LHC with
    $\sqrt{s}=14\TeV$.}
  \label{fig:phythia3levels}
\end{figure}

These predictions are valid at parton level only, so it is interesting
to investigate the impact of hadronisation and underlying event on
event-shape distributions. This can be done with MC event
generators. Figure~\ref{fig:phythia3levels} shows that $k_t$-algorithm
jet resolution parameters, for example $y_{3,g}$, are essentially
not affected by hadronisation and underlying event, while event
shapes, like $\rho_{T,{\cal E}}$, get moderate hadronisation
corrections, falling as an inverse power of the jet $p_t$, but get a
huge contribution from the underlying event. This different
sensitivity shows that event-shape distributions can be exploited for
the validation of MC event generators. Jet resolution parameters are
better suited for tunings of parton shower parameters, while with
remaining event shapes one can test models of hadronisation and
underlying event.

\section{Event shapes for New Physics}
\label{sec:evshp-NP}

\begin{figure}[h]
  \centering
  \includegraphics[width=.25\textwidth]{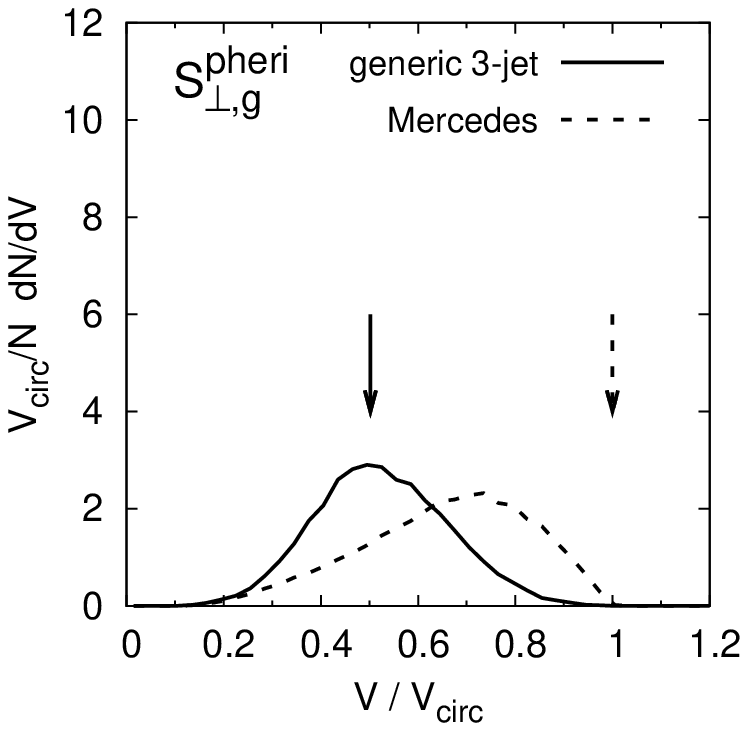}
  \rule{.03\textwidth}{0cm}
  \includegraphics[width=.25\textwidth]{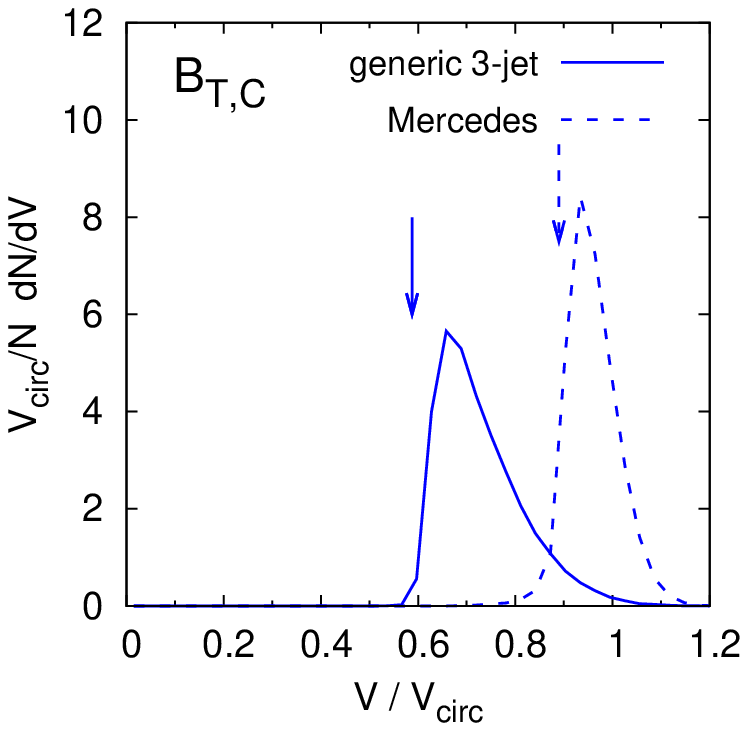}
  \rule{.03\textwidth}{0cm}
  \includegraphics[width=.25\textwidth]{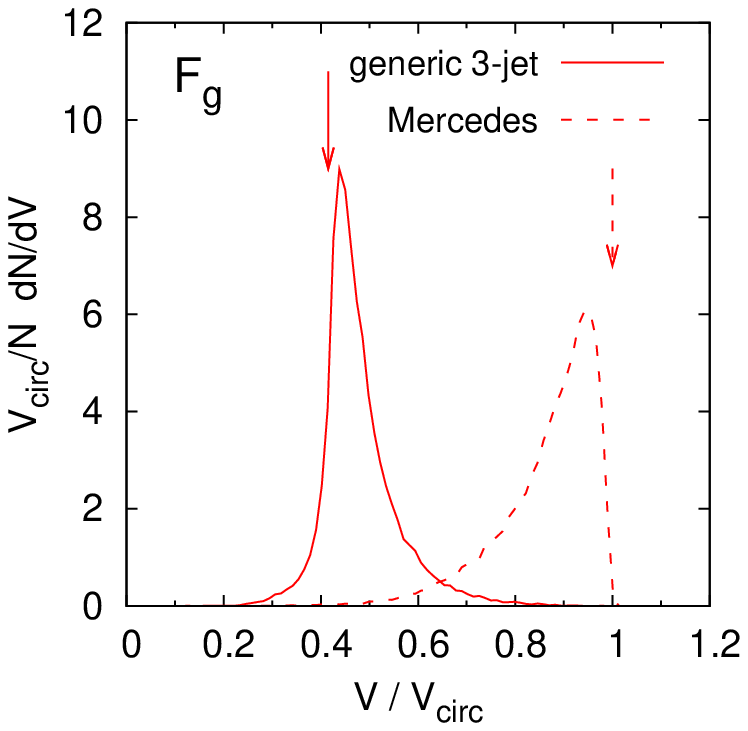}
  \caption{Distribution in the value of three different variables
    obtained from events with two different hard configurations
    dressed with parton shower. The two starting configurations are a
    generic three-parton and a totally symmetric (Mercedes) event in
    the transverse plane.}
  \label{fig:spheri-broad}
\end{figure}

A common use of event shapes is that of discriminating among events
with different topologies. This is particularly important in New
Physics searches, where one expects events with massive particles to
be much broader than dijet events. We have then tried to assess the
performance of known hadronic event shapes for such studies. First,
when considering symmetric events with an arbitrary number of
particles in the transverse plane, one can only distinguish between
two- and multi-jet events, irrespectively of the number of jets. One
can then try to discriminate among different topologies in a sample
with the same number of jets (three jets in the considered case). One
then finds (see figure~\ref{fig:spheri-broad}) that infrared and
collinear safe variables fare much better in this respect than unsafe
ones (like the widely used transverse sphericity).
We remark also that event shapes like the broadenings, which treat
tranverse and longitudinal degrees of freedom on equal footing, are
better suited for the identification of massive particle decays, since
their value is hardly affected by the orientation of the event plane.

For practical applications, it is however desirable to have variables
that are more sensitive to the spherical limit. One example presented
in~\cite{Banfi:2010xy} is the {\em supersphero} observable which is
non-zero only for events in which there are three non-coplanar
particles in each of the ``hemispheres'' in which the event is divided
by the transverse thrust axis.  We believe that phenomenological
applications of variables like supersphero, as well as better
final-state observables for New Physics, constitute an important
subject that deserves further studies.

\end{document}